\documentclass{emulateapj}
\usepackage{natbib}
\usepackage{hyperref,verbatim}

\newcommand{\sigmaDM}{{\sigma_{\textsc{\tiny DM}}/m}}

\begin{document}

\title{The Mismeasure of Mergers: Revised Limits on Self-interacting
  Dark Matter in Merging Galaxy Clusters}

\shorttitle{Dark Matter in Merging Galaxy Clusters}

\author{David Wittman\altaffilmark{1,2}, Nathan Golovich\altaffilmark{1},
  William A. Dawson\altaffilmark{3}}

\altaffiltext{1}{Physics Department, University of California, Davis,
  CA 95616; dwittman@physics.ucdavis.edu}

\altaffiltext{2}{Institito de Astrof\'{i}sica e Ci\^{e}ncias do
  Espa\c{c}o, Universidade de Lisboa, Lisbon, Portugal}

\altaffiltext{3}{Lawrence Livermore National Laboratory, P.O. Box 808
  L-210, Livermore, CA, 94551, USA}


\begin{abstract} 
  In an influential recent paper, \citet{Harvey15} derive an upper
  limit to the self-interaction cross section of dark matter
  ($\sigmaDM < 0.47$ cm$^2$/g at 95\% confidence) by averaging the
  dark matter-galaxy offsets in a sample of merging galaxy clusters.
  Using much more comprehensive data on the same clusters, we identify
  several substantial errors in their offset measurements.  Correcting
  these errors relaxes the upper limit on $\sigmaDM$ to $\lesssim 2$
  cm$^2$/g, following the \citet{Harvey15} prescription for relating
  offsets to cross sections in a simple solid body scattering
    model. Furthermore, many clusters in the sample violate the
  assumptions behind this prescription, so even this revised upper
  limit should be used with caution.  Although this particular sample
  does not tightly constrain self-interacting dark matter models when
  analyzed this way, we discuss how merger ensembles may be used more
  effectively in the future.  We conclude that errors inherent in
    using single-band imaging to identify mass and light peaks do not
    necessarily average out in a sample of this size, particularly
    when a handful of substructures constitute a majority of the
    weight in the ensemble.
\end{abstract}

\section{Introduction}

Dark matter (DM) comprises most of the matter in the universe but
little is known about its properties.  It has been detected
gravitationally, but despite many searches there is as yet no evidence
it participates in other known interactions.  Perhaps the dominant DM
particle model has been the weakly interacting massive particle
(WIMP), which is difficult to detect directly.  Searches for weak
interactions of dark matter with normal matter have rapidly improved
in sensitivity, however, and are beginning to rule out interestingly
large regions of parameter space \citep[see][for a review]{Klasen15}.
Another class of models, hidden sector models, posit substantial
interactions {\it between} DM particles even if nongravitational
interactions with normal matter are undetectably weak or nonexistent
\citep{feldman2007,feng2008,cohen2010}.  Empirical constraints on DM
self-interactions rely on astrophysical arguments and are much less
stringent than those on interactions between DM and normal matter.
Upper limits from astrophysical arguments \citep[e.g.,][]{Randall2008}
are on the order of one cm$^2$/g, or two barns per GeV---twenty orders
of magnitude larger than the cross-sections relevant to collider and
direct-detection searches.
\footnote{DM cross sections are cited in terms of area per unit mass
  because the rather than area per particle because astronomers are
  able to measure the total mass of a collection of DM particles but
  not the mass per particle.}

In other words, if DM particles interact with each other with about
the same cross section as neutrons do, this interaction could still
have escaped detection to date.  This has provided general motivation
for efforts to tighten astrophysical constraints on $\sigmaDM$.  A
more specific motivation comes from measurements of dwarf galaxy and
galaxy cluster density profiles, which are suggestive of
self-interacting DM (SIDM) with a cross section around 0.1--5 cm$^2$/g
\citep{BoylanKolchin12,Rocha2012,Peter2012,Sand08,Newman2012a,Newman2012b,Elbert15}.
If true, this would exclude the WIMP model by some twenty orders of
magnitude and point the way to some truly new physics.  Skeptics,
however, point out that these astrophysical environments also contain
difficult-to-model baryonic effects that could explain the measured
profiles \citep[e.g.][]{Brooks14}.

A complementary method for probing some types of self-interactions may
be to analyze mergers of galaxy clusters (which are mostly DM by mass)
as ``dark matter colliders.''  The well-known Bullet Cluster serves as
the best example and yields, at 68\% confidence, $\sigmaDM < 0.7$
cm$^2$/g based on a mass-to-light argument and $\sigmaDM < 1.25$
cm$^2$/g based on the offset argument described in more detail below
\citep{Randall2008}.  An ensemble of such mergers could potentially
drive this upper limit down enough to confirm SIDM if it exists.
Because cluster mergers and galaxy cores probe different velocity
scales, such a detection would also characterize the velocity
dependence of the interaction and thereby constrain the mediator mass
\citep{Loeb11,Zavala13,Manoj16}.  Even if one prefers to view this
work as an exercise in excluding SIDM, observations on both low and
high velocity scales will be necessary to impinge on the broad class
of SIDM models that are naturally velocity dependent \citep{Loeb11}.

Hence there is great interest in the result of \citet[hereafter
H15]{Harvey15}, who used offsets between galaxies and DM in an
ensemble of 30 merging clusters to derive, for a simple solid
  body scattering model, an upper limit of $\sigmaDM <0.47$ cm$^2$/g
at 95\% confidence and $<-0.01$ cm$^2$/g at 68\%
confidence.\footnote{The 68\% confidence upper limit is not stated
  directly by H15 but is implied by their Figure 4.} This is a
dramatic improvement on the previous best constraint from offsets
cited above, and thus has already helped drive new constraints on the
velocity dependence and therefore the mediator mass \citep{Manoj16}.
Because this dramatic improvement has profound implications for
particle models, it warrants further scrutiny. Many of the merging
clusters in the H15 sample have been intensively studied individually,
thus providing independent measurements of numerous DM-galaxy offsets.
These independent measurements derive from heterogeneous data sources
and analysis methods, but always involve {\it more} data and analysis
than H15 applied to any individual cluster. In this paper we use the
more extensive data to reveal substantial errors that, when corrected,
greatly loosen the H15 constraint.

The remainder of the paper is organized as follows.
Section~\ref{sec-technique} reviews the basic premise behind the
merging cluster technique and outlines the H15 procedure in enough
detail to understand which substructures are most highly weighted in
the final result.  In Section~\ref{sec-lit} we review the literature
on these highly-weighted substructures and either accept the H15
measurement, improve the H15 measurement, or argue that the
substructure is unusable for this test.  In Section~\ref{sec-results}
we analyze the updated catalog using the H15 formalism and derive
updated SIDM constraints.  In Section~\ref{sec-discuss} we discuss the
result in the broader context of astrophysical tests of dark matter,
and present some concluding remarks.

\section{The Merging Cluster Technique}\label{sec-technique}

\begin{figure}
\centerline{\includegraphics[scale=0.22]{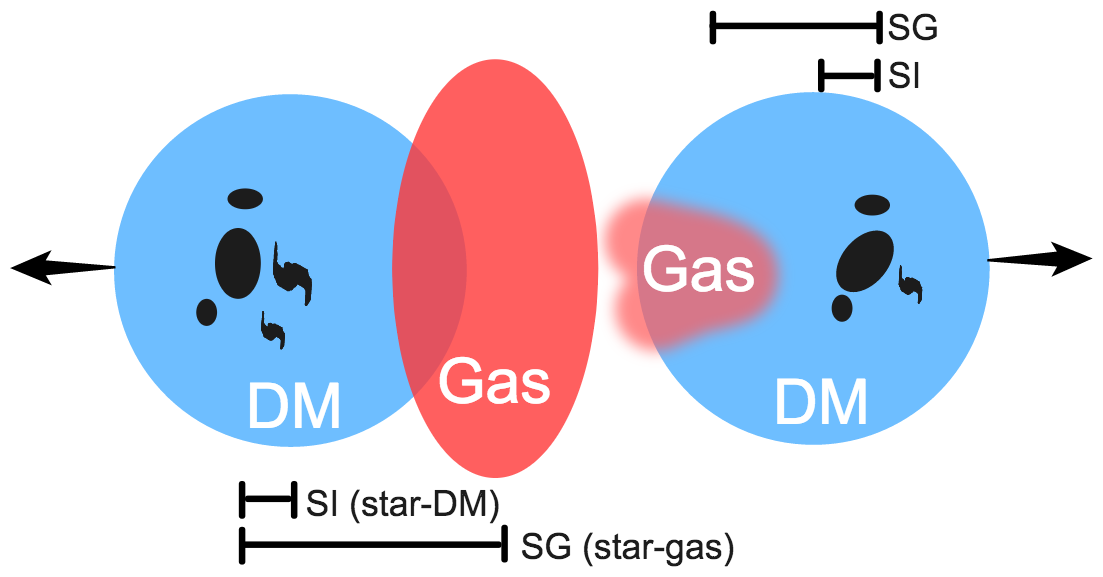}}
\caption{Schematic merger scenario: two subclusters have passed
  through each other, and the gas associated with each has slowed due
  to momentum exchange. This is observable as an offset between the
  star (i.e., galaxy) and gas positions, $\delta_{SG}$.  In analogy,
  any star-DM offset $\delta_{SI}$ may be attributed to momentum
  exchange between the DM halos and thus related to a cross section
  $\sigmaDM$.  Subcluster masses and gas densities may vary
  considerably.}\label{fig-premise}
\end{figure}

A galaxy cluster consists of gas, DM (constituting the great majority
of the mass), and galaxies.  When two such clusters fall together, the
two sets of galaxies pass through each other with little or no
exchange of momentum.  The gas clouds, in contrast, exchange momentum
and thereby slow down compared to the galaxies.  A snapshot of a
system soon after pericenter passage, for example the well-known
Bullet Cluster, shows the two gas clouds closer to the center of the
combined system and the galaxies farther out.  The (at least
approximately) collisionless nature of dark matter is then
demonstrated when gravitational lensing shows that the majority of the
mass (and by implication the DM) is coincident with the galaxies
rather than the gas \citep{Markevitch04}. If DM in fact exchanges some
momentum in a way analogous to the gas, the DM at this stage of the
merger will be located between the galaxies and the gas
(Figure~\ref{fig-premise}).  The observed DM location thus constrains
the DM self-interaction cross-section $\sigmaDM$ in this
model.\footnote{Other self-interaction models are not well probed by
  the DM offset but may be probed with other observations; see
  Section~\ref{sec-discuss}.}

The same reasoning applies to infall of smaller structures.  In the
limit of small structures falling into a much more massive structure,
\citet{Harvey14} developed an analytical relation between $\sigmaDM$
and the galaxy-DM-gas geometry.  Defining the galaxy-gas separation as
$\delta_{SG}$ ($S$ stands for ``star'' which is synonymous with
``galaxy'' in this context), they define a coordinate system starting
at the galaxy location and stretching to the gas location.  The key
observable is the DM displacement along this coordinate system,
$\delta_{SI}$, in units of $\delta_{SG}$.  This ratio,
${\delta_{SI}\over \delta_{SG}}\equiv \beta$, has a simple analytical
relationship to $\sigmaDM$ if DM behaves analogously to the gas (see,
however, Section~\ref{sec-discuss} for caveats on this analogy).
$\beta=0$ corresponds to collisionless DM, $\beta=1$ to DM just like
baryonic gas, and intermediate values correspond to intermediate
cross-sections.  H15 averaged over 72 substructures and found
$\langle\beta\rangle = -0.04\pm 0.07$; negative values are unphysical
but indicate that the data are in tension with the idea of momentum
transfer between the DM halos.

H15 analyzed 72 substructures in 30 systems.  To identify the
substructures with greatest influence on the ensemble result, note
that standard propagation of errors on the ratio $\beta\equiv
{\delta_{SI}\over \delta_{SG}}$ yields
\begin{equation}
\sigma^2_\beta = {\sigma_{SI}^2\over \delta_{SG}^2} + {\sigma_{SG}^2 \delta_{SI}^2\over \delta_{SG}^4}
\end{equation} 
H15 adopt an uncertainty of $\sigma_{SG}=\sigma_{SI}=60$ kpc on each
offset measurement.  Therefore we can factor this out and write
\begin{equation}
\sigma^2_\beta \propto {1\over \delta_{SG}^2}(1 + {\delta_{SI}^2\over \delta_{SG}^2}). 
\end{equation}
With inverse-variance weighting, then, the weight of the $i$th
substructure would be
\begin{equation}
w_i \propto {\delta_{SG,i}^2\over 1 + \delta_{SI,i}^2/ \delta_{SG,i}^2}.\label{eqn-3}
\end{equation}
Because $\delta_{SI,i}^2/ \delta_{SG,i}^2 <<1$ in most cases,
$w_i \propto \delta_{SG,i}^2$ is a good approximation for quick
assessment of the importance of a particular substructure in the
ensemble.  Although H15 multiply Gaussian probability density
functions (PDFs) rather than compute a single inverse-variance
weighted mean, the effect is the same: substructures with large
$\delta_{SG,i}$ predominantly determine the result.  This makes
intuitive sense, because the ratio
$\beta\equiv {\delta_{SI}\over \delta_{SG}}$ is highly uncertain when
the denominator is small compared to its 60 kpc uncertainty.
Conversely, a large $\delta_{SG,i}$ provides a stable baseline from
which to measure the $i$th ratio, and this results in a narrow PDF, or
effectively a large weight for the $i$th substructure.

Of course, one may question the adoption of
$\sigma_{SG}=\sigma_{SI}=60$ kpc for each offset measurement, because
the accuracy of offset measurements may vary substantially from
substructure to substructure.  Our immediate goal is to identify the
substructures with the most influence on the H15 result, so we defer
discussion of this point to Section~\ref{sec-discuss}.

Table~\ref{tab-changes} in Section~\ref{subsec-summary} lists, in
decreasing order, the weight of each substructure as a percentage of
the total weight.  The list is truncated after 16 substructures
comprising 85\% of the total weight.  Next, we perform a literature
review of these 16 substructures in descending order of weight.

\section{Literature Review}\label{sec-lit}

We believe this paper will be most useful to the community if it
highlights a handful of substantial inaccuracies in H15, rather than
revisiting every detail.  The following review thus defaults to
respecting each H15 measurement unless the literature provides strong
evidence to the contrary. Given the heterogeneity of the data, what
constitutes ``strong evidence'' may vary.  While acknowledging that
this approach could lead to bias (discussed further in
Section~\ref{sec-discuss}), we are confident that readers will agree
with our corrections in most cases. In addition to outright
corrections, we will discard substructures for which the matching
between gas, DM, and galaxy components is uncertain (e.g.,
Subsection\ref{subsec-A2744}).  In principle, one may choose instead
to model such cases; heavy tails could reflect the probability that
the given gas, DM, and galaxy components were never coincident in the
past.  However, quantifying this probability would be very difficult.
Furthermore, we suspect that substructures with such heavy-tailed PDFs
would contribute very little to an ensemble constraint.  We therefore
simply discard such cases.

Throughout this review, keep in mind that H15 used only single-band
imaging (not necessarily the same band for each cluster) in their main
analysis in order to yield a large and (in some respects) homogeneous
sample.  Multiband data allow for better selection of lensing sources
and better exclusion of foreground galaxies when mapping the light
distribution.  Multiband imaging and spectra are also necessary to
support a strong lensing analysis, which can locate the mass much more
precisely than a weak lensing analysis.  In fact, each of the highly
weighted systems identified in Table~\ref{tab-changes} has been
studied in more detail with some combination of these techniques.  The
bands observed, the availability of spectroscopy and strong lensing
information, and the data processing and analysis choices vary.
Nevertheless, we believe it would be wrong to ignore studies that
employ far more data and more robust methods than H15 do on the very
same merging systems.

Despite the heterogeneity, a few general remarks do apply.  First, we
do not seek to update the gas positions, because those are unaffected
by the use of the additional data and techniques listed above.  Of
course X-ray analysis choices such as point source removal and
smoothing scale are important, but this paper focuses on what can be
learned from {\it additional data}.  Second, we generally keep the
nominal H15 uncertainty of 60 kpc on each offset, because the papers
we draw from generally do not offer a detailed uncertainty analysis on
these particular quantities.  Third, lensing is sensitive to all forms
of mass, not just DM, so the lensing position must be corrected for
the gas mass contribution to obtain a position for the DM alone.
Papers that supply more accurate lensing positions usually do not
supply information necessary to make this correction, so---except in
cases where more specific information is available---we adopt the mean
H15 correction of $-4.3\pm 1.6$ kpc (as a reminder, this is in a
coordinate system originating at the galaxy position and increasing
toward the gas position).  The 1.6 kpc uncertainty in the mean of 72
substructures suggests a sample standard deviation of 13.6 kpc. For
most substructures the nominal uncertainty in each offset is 60 kpc so
the uncertainty in the gas mass correction is highly subdominant and
will be neglected unless otherwise specified.

\subsection{Abell 2744 (Northwest)\label{subsec-A2744}} 

\begin{figure}
\centerline{\includegraphics[scale=0.35]{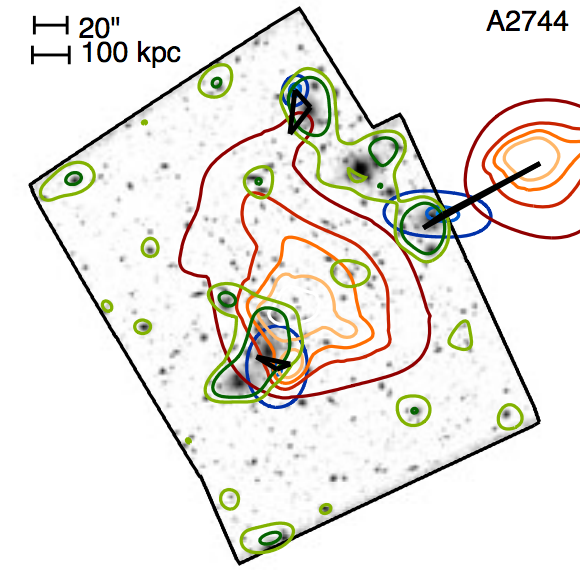}}
\centerline{\includegraphics[scale=0.5]{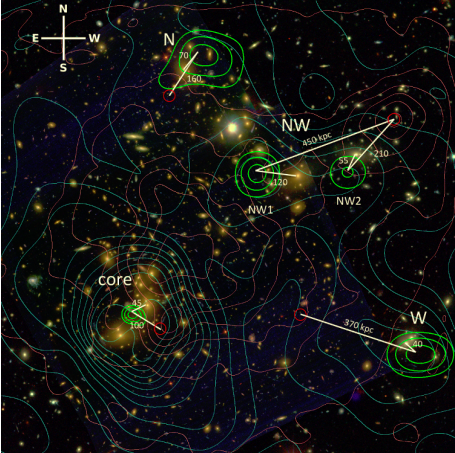}}
\caption{{\it Top:} View of Abell 2744 from H15. In all panels from
  H15, X-ray (gas) contours are reddish, green contours indicate
  visible light, and blue contours indicate mass inferred from
  gravitational lensing.  In the highly weighted western subcluster
  the heavy black line indicates nearly collinear alignment of
  galaxies, DM and gas.  {\it Bottom:} map from the more detailed
  analysis of \citet{Merten11}, with x-ray (red), lensing (cyan), and
  confidence contours for lensing peak locations (green, with
  $0.3\sigma, 1\sigma, 2\sigma$ contours). The H15 mass peak is now
  resolved into two peaks (NW1 and NW2), contradicting the H15
  assumption of simple infall of a gas-DM-galaxy
  substructure.}\label{fig-A2744}
\end{figure}

The top panel of Figure~\ref{fig-A2744} shows the H15 map of this
cluster.  To orient the reader, each H15 panel portrays one merging
system, and each system has at least two substructures.  Therefore, at
least two independent offsets can be measured from each system (three
in this case).  In each H15 panel, the red contours indicate the
surface brightness of the hot X-ray emitting gas, the green contours
indicate galaxy brightness, and the blue contours indicate the mass
(primarily DM) distribution as inferred from weak gravitational
lensing.  H15 draw a triangle connecting the peaks of the three
distributions (gas, galaxies, mass) in each subcluster.  If DM
exhibits a drag force, we expect to find the mass peak between the
galaxies and the gas, yielding $\delta_{SI}>0$. This is referred to as
DM ``lagging'' the galaxies because the gas definitely lags the
galaxies (until turnaround; see Section~\ref{sec-discuss}).  Lateral
displacements are considered irrelevant---resulting from measurement
error and perhaps other stochastic processes---so $\delta_{SI}$ is
actually the projection of the galaxy-DM leg of the triangle onto the
galaxy-gas leg.

The length of the galaxy-gas leg, $\delta_{SG}$, determines the
importance of the substructure in the ensemble analysis
(\S\ref{sec-technique}); the top panel of Figure~\ref{fig-A2744} shows
that the northwest\footnote{H15 and this paper follow the astronomical
  convention of placing north up and east left on the page.}
substructure is by far the most important.  This substructure is the
most highly weighted in the entire ensemble, with 17\% of the total
weight, and we focus on it exclusively in this subsection.  The
corresponding H15 triangle appears to be a long line segment due to
negligible lateral displacement.  This triangle extends off the HST
field of view because the Chandra X-ray Observatory, used to locate
the gas, has a much larger field.  The galaxy and DM components of
this substructure are located on the edge of the HST field, with the
DM trailing the galaxies, as predicted by the drag-force model if
$\sigmaDM>0$.

\citet{Merten11} performed a detailed strong lensing, weak lensing and
X-ray analysis of this cluster, imaging a larger area so the relevant
substructure is no longer on the edge of the optical field (bottom
panel of Figure~\ref{fig-A2744}).  They supplemented two-band HST
imaging with ground-based VLT and Subaru imaging, as well as 118
spectroscopic redshifts to guide the photometric selection of source
galaxies for the lensing analysis.  They resolved the western mass
peak found by H15 into two distinct mass peaks (labeled NW1 and NW2 in
the lower panel of Figure~\ref{fig-A2744}), separated by $\sim 200$
kpc but on the {\it same} side of the gas.  \citet{Merten11} link both
mass peaks to the gas peak via a complicated scenario far outside the
\citet{Harvey14} framework of an infalling subcluster experiencing a
small separation between its gas, DM, and galaxy
components. \citet{Jauzac2016} developed a detailed strong-lensing
mass model using additional data and confirmed the existence of NW1
and NW2, but found them to be coincident with their respective nearby
bright galaxies. Meanwhile, \citet{Medezinski2016} analyzed Subaru and
VLT imaging and derived quite different locations for the DM
substructures in this system, as well as a different merger scenario.
Given these complicated and competing merger scenarios, we cannot be
certain that NW1 and/or NW2 were ever united with this particular gas
peak.  With the gas-galaxy-DM association itself in question, the
offsets are not meaningful and the substructure should be omitted from
the ensemble. In fact, \citet{Harvey14} discussed association
uncertainty and stated, correctly, that such uncertainty would vanish
for substructures with offsets small enough to satisfy their
approximation ($<30$ kpc). The association uncertainty arises here
because $\delta_{SG}\sim 400$ kpc, the largest in the H15
ensemble. Even if more robust associations can be made in the future,
this substructure will never satisfy the \citet{Harvey14}
approximation.

Discarding this substructure should {\it strengthen} the H15 case
against SIDM, because this is a very highly weighted substructure
that---contrary to the H15 ensemble overall---does have a DM offset in
the direction predicted by SIDM.  It is worth noting, for illustration
purposes only, that the configuration of green, blue, and red contours
representing this substructure in the top panel of
Figure~\ref{fig-A2744} is to be expected if $\sigmaDM \approx 1$
cm$^2$/g.  Specifically, for this cross section Equation~1 of H15,
predicts $\beta\equiv\frac{\delta_{SI}}{\delta_{SG}}=0.14$, while the
(discarded) value here is 0.19. The substructures in H15 more
typically exhibit green (galaxy) contours {\it between} the blue (DM)
and red (gas) contours, corresponding to a negative $\beta_i$ and a
negative (unphysical) cross section.

In the remaining subsections of the literature review we will not
comment on the impact of each particular correction to the H15
catalog; the foregoing explanation should enable the reader to do so
if desired.  For Abell 2744, we reiterate that omitting this
substructure does nothing to loosen the constraints on
$\sigmaDM$, quite the opposite in fact.  Our final conclusion
that the ensemble constraints are quite loose will owe nothing to the
omission of this particular substructure.


\subsection{DLSCL J0916.2+2951 (South)\label{ssec-musket}}

\begin{figure}
\centerline{\includegraphics[scale=0.5]{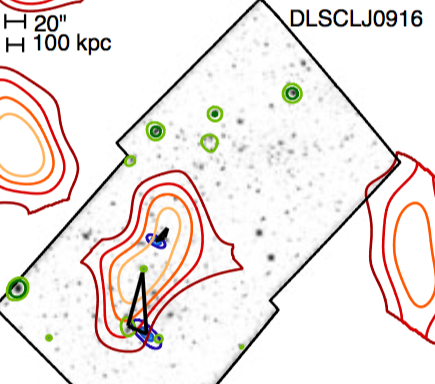}}
\centerline{\includegraphics[scale=0.35]{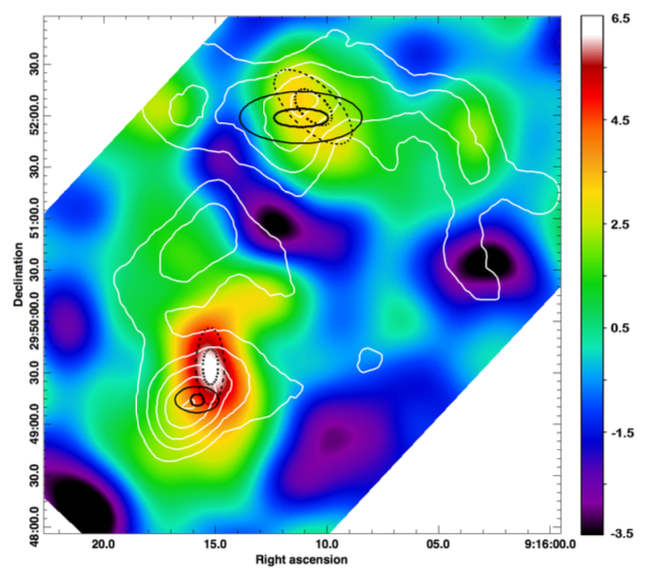}}
\caption{{\it Top:} View of DLSCL J0916 from H15. In the highly
  weighted southern subcluster, H15 find the mass to be ahead of the
  galaxies ($\delta_{SI} = -19$ kpc), corresponding to a negative
  $\sigmaDM$. {\it Bottom:} the more detailed analysis of
  \citet{DawsonPhDT}, with mass in colorscale and galaxy luminosity
  density in white contours. In the south, the galaxy position agrees
  with that of H15 and thus serves as a reference point for comparing
  the two panels; \citet{DawsonPhDT} find the mass to be trailing the
  galaxies.  Solid (dashed) ellipses are 68\% and 95\% confidence
  intervals for the galaxy luminosity (mass) centroid.}
\label{fig-musket}
\end{figure}

\citet{Dawson11} and \citet{DawsonPhDT} studied this cluster with two
bands of HST/ACS imaging, five bands of deep ground-based imaging, and
634 spectroscopic redshifts to support the background source and
galaxy member selection. Figure~\ref{fig-musket} compares the
\citet{DawsonPhDT} result with that of H15.  The southern triangle in
the top panel of Figure~\ref{fig-musket} illustrates the H15 finding
that the mass is actually ahead of the galaxies; H15 found
$\delta_{SI} = -19$ kpc.  The lower panel, from \citet{DawsonPhDT}, shows the
lensing mass in a colorscale with galaxy luminosity density overlaid
in white contours.  The mass is clearly lagging the galaxies;
\citet{DawsonPhDT} found the offset to be $+129$ kpc based on the
lensing peak alone, and $+80$ kpc after modeling out the gas mass to
yield the DM mass alone.

While the H15 uncertainties are large enough to encompass the
\citet{DawsonPhDT} value, the H15 mass position is outside the
\citet{DawsonPhDT} $2\sigma$ confidence ellipse (the larger dashed
ellipse).  Because the \citet{DawsonPhDT} value is supported by
multiband selection of lensing sources and member galaxies, backed up
by extensive spectroscopy, we adopt $\delta_{SI}=80$ kpc.


\subsection{Abell 520}

\begin{figure}
\centerline{\includegraphics[scale=0.4]{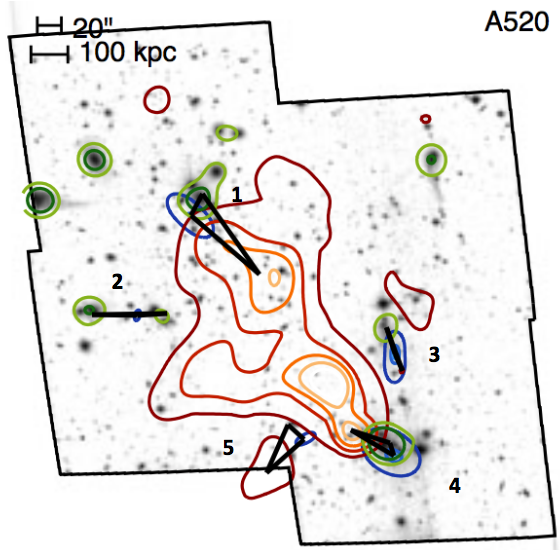}}
\centerline{\includegraphics[scale=0.35]{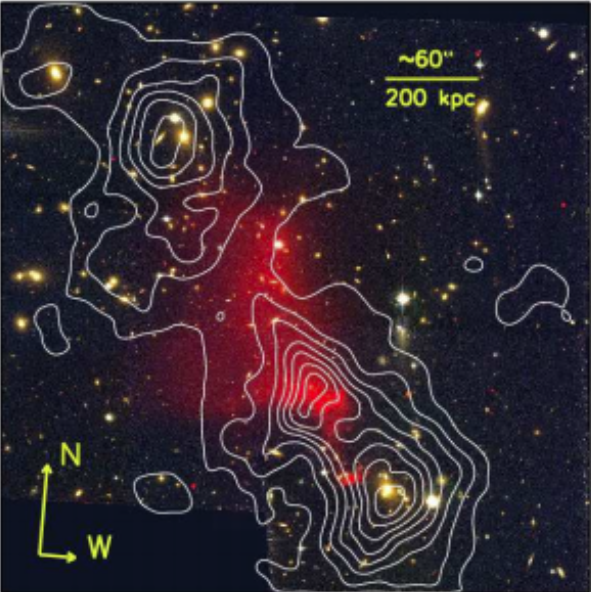}}
\centerline{\includegraphics[scale=0.55]{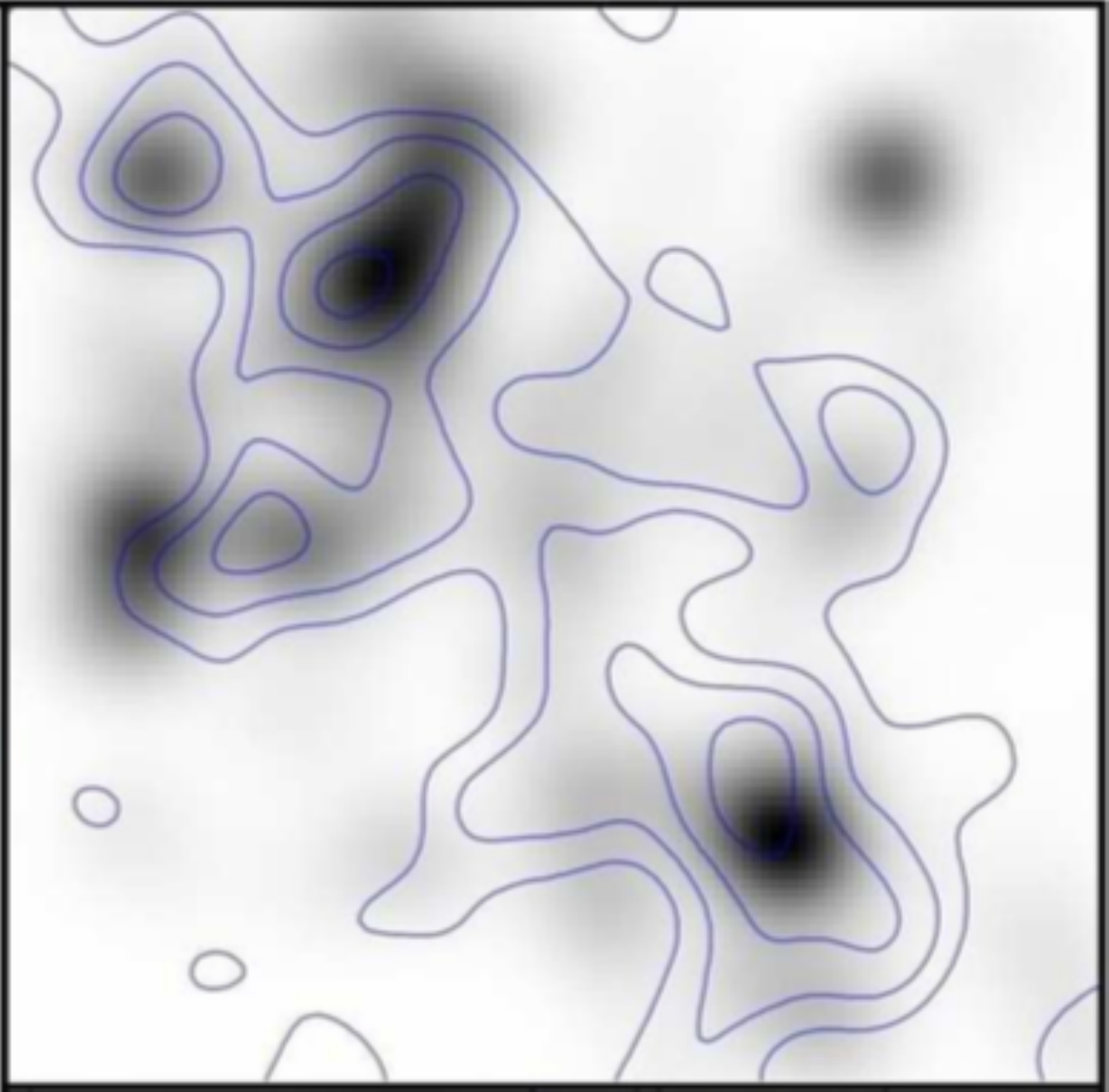}}
\caption{{\it Top:} H15 analysis of Abell 520.  The numerical labels
  are our annotation, to clearly link to descriptions of substructures
  in the text. {\it Middle:} multiband lensing analysis from
  \citet{Jee14-A520}, with X-ray in redscale and lensing contours in
  white.  {\it Bottom:} multiband lensing analysis from
  \citet{Clowe12-A520} (contours) on top of smoothed galaxy light (grayscale).}\label{fig-A520}
\end{figure}

All five of the Abell 520 substructures identified by H15 are in the
top 16 substructures by weight, so we review the entire system at
once.  In the top panel of Figure~\ref{fig-A520} we label the H15
substructures for reference below; the middle and bottom panels
  show multiband lensing analyses from \citet{Jee14-A520} and
  \citet{Clowe12-A520} respectively.  The latter two analyses broadly
  agree, with clear agreement on the main peaks (labeled 1 and 4 here)
  and both maps showing extensions toward the H15 peak labeled 5.  The
  two maps agree weakly on the H15 peak we label 2, which appears as a
  peak in \citet{Clowe12-A520} but as a weaker extension of contours
  in \citet{Jee14-A520}. For our purposes, the major disagreement
  between the two detailed lensing analyses of Abell 520 is that
  \citet{Clowe12-A520} find a west-central peak, in the rough
  vicinity of H15 peak 3, while \citet{Jee14-A520} does not.


{\it Substructure 1:} this northern subcluster is the most highly
weighted of the five, with about 10\% of the total weight of the H15
sample.  This substructure is consistently identified and located by
\citet{Jee14-A520} and \citet{Clowe12-A520}.  Their maps qualitatively
agree with H15, so we adopt the H15 offset.

{\it Substructure 2:} this is the second most highly weighted
substructure in the system, with about 5\% of the total weight of the
H15 sample.  However, there is no gas peak in the area.  The absence
of a gas peak prevents us from defining any gas-star-DM geometry. We
therefore recommend omitting this putative substructure from the
catalog.

We will see several more examples where the X-ray peak is not apparent
in the H15 figure, so we address the issue more generally
here. Presumably the gas peak in this substructure is not absent but
merely at too low a signal-to-noise ratio (S/N) to appear on the H15
panel, which shows only a handful of contour levels. However, we see
no indication of a diffuse X-ray source at this location in other
presentations of the X-ray data, for example the pixelized redscale in
the \citet{Jee14-A520} panel, or the detailed X-ray analysis of this
system presented by \citet{Markevitch05}. H15 automatically matched
gas, galaxy, and DM peaks by searching for the nearest peak, so we
suggest that this algorithm simply found a very low S/N local X-ray
maximum.  Many such local maxima must exist due to photon noise and
unidentified point sources, but they should not be used to define
substructures and offsets.

{\it Substructure 3:} the ``gas peak'' again appears to be a very low
S/N local maximum with minimal anular extent. Furthermore, the H15
mass peak does not correspond to a peak in \citet{Jee14-A520}. 
  \citet{Clowe12-A520} did find a center-west mass peak, but it lies
  270 kpc from the H15 location, on the opposite side of a neaby set
  of galaxies. Finally, most of the light in the green H15 contour is
actually a streak from the bright star to the north rather than galaxy
light---this is more easily seen by comparison to the middle panel of
Figure~\ref{fig-A520}, which has fewer overlays. The streak is
probably caused by some combination of CCD bleeding, charge transfer
inefficiency, and the star being near the chip gap, which makes its
effects on the neighboring chip more difficult to remove.  The
appearance of this streak thus varies greatly depending on the
specifics of the image processing, for example it is absent in most
press release images. Note also that \citet{Clowe12-A520} used
  color information to select member galaxy light, and their
  luminosity map (grayscale in the bottom panel of
  Figure~\ref{fig-A520}) is extremely faint in this area.  H15 kindly
allowed us to inspect a higher-resolution version of their image, and
the streak begins further north and contains more total light than in
the \citet{Jee14-A520} image.  Without the streak, it is unclear where
the galaxy luminosity peak would be located.

In summary, the literature confirms neither the gas, the galaxy
luminosity, nor the lensing mass peak for this substructure, so we
remove it from the ensemble analysis.

{\it Substructure 4:} along with Substructure 1, this is one of the
two widely confirmed substructures in the system, and the H15
locations are consistent with previous work in the literature.  We
adopt the H15 offset.  Note that this substructure carries
substantially less weight than the northern substructure---at 1.6\% of
the total weight, it is the least weighty substructure in our
literature review.

{\it Substructure 5:} any X-ray emission in this area is too faint to
be seen in the \citet{Jee14-A520} panel of Figure~\ref{fig-A520}.  The
less highly processed X-ray image of \citet{Markevitch05} shows what
is possibly a local maximum at this location, but it is difficult to
argue that it represents a separately identifiable gas concentration.
Furthermore, the visible-light peak in the H15 panel clearly consists
of a single galaxy.  The \citet{Jee14-A520} and
  \citet{Clowe12-A520} mass contours do extend into this general area,
  but the H15 lensing peak lies about 100 kpc from the ``spine'' of
  those extensions.  Lacking a clear association of gas, galaxy, and
  DM peaks, we recommend removing this putative substructure from the
ensemble analysis.


\subsection{1E0657-56 (the Bullet Cluster)}

Both substructures in this system are highly weighted in the H15
analysis: the eastern (main) subcluster ranks fourth with with 6.1\%
of the total weight, and the western (bullet) subcluster ranks ninth
with 3.8\% of the total weight. These weights reflect how well the gas
and galaxies are separated in this system, clearly establishing the
baseline to which the galaxy-DM separation must be compared.  This
cluster has been extensively studied with lensing, beginning with
  the weak lensing analysis of \citet{Clowe04}, who used imaging from
  the Very Large Telescope, with two bands ($B$ and $I$) for color
  selection of source galaxies. Another weak lensing analysis,
  \citet{Clowe06}, is based on a wide array of independent data: $BVR$
  imaging from the ESO 2.2m telescope, $BVR$ imaging from the Magellan
  6.5m telescope, and F606W imaging from HST/ACS, plus F435W and F814W
  imaging on the western subcluster.  \citet{Bradac09-Bullet} used
  that data plus additional HST data to produce a combined strong and
  weak lensing analysis.  More recently, \citet{Paraficz16} produced a
  strong-lensing mass map using 14 multiply-imaged systems.

\begin{figure}
\centerline{\includegraphics[scale=0.5]{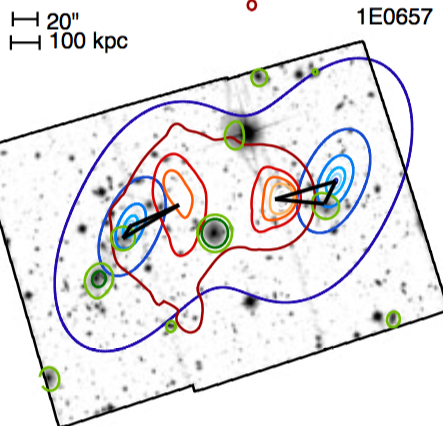}}
\centerline{\includegraphics[scale=3.3]{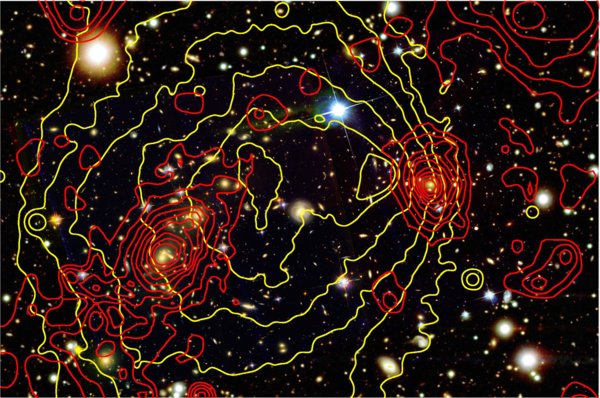}}
\caption{{\it Top:} View of the Bullet Cluster from H15.  {\it
    Bottom:} maps from the more detailed analysis of
  \citet{Bradac09-Bullet}, with mass contours from strong and weak
  lensing in red, and X-ray in yellow.  The H15 lensing position for
  the western subcluster, putting the mass ahead of the galaxies, is
  highly excluded based on this more detailed
  analysis. }\label{fig-bullet}
\end{figure}

We show the \citet{Bradac09-Bullet} mass map
  (Figure~\ref{fig-bullet}) because it gives the reader a sense of
  both weak and strong lensing smoothing scales.  Most of the area
  shown is constrained only by weak lensing, so the lower contours
  illustrate the smoothing and noise typical of weak lensing, while
  strong lensing guides some of the details in the higher contour
  levels. We stress that regardless of the specific reconstruction
  technique, all the above-cited analyses are in good agreement, and
  none show a displacement between mass and galaxy light in either
  substructure.

{\it East:}  the H15 lensing contours place the mass about 130 kpc
  northwest of its position in the above-cited lensing analyses. Given
  the agreement among the above-cited sources, their strong exclusion
  of the H15 position (for example, the H15 position is well outside
  the \citet{Clowe06} 99.7\% confidence interval), and their
  advantages in data quality (such as multiple bands for source
  selection), we argue that the H15 position is not credible.
  However, it would be misleading to use the consensus lensing
  position with the H15 light position. This is because H15 split the
  light into two separate peaks and chose the lesser one because of
  its proximity to their lensing peak. When the mass and light are
  smoothed on similar scales as in \citet{Clowe04}, there is no
  discernible difference between the mass and light positions. We
  therefore adopt an offset of zero.

{\it West:}  \citet{Clowe04},
\citet{Clowe06}, \citet{Bradac09-Bullet}, and \citet{Paraficz16} all
disagree with H15 and agree with each other in placing the mass much
closer to the light. The H15 position is well outside the
\citet{Clowe06} 99.7\% confidence interval and outside most of the
contours in the \citet{Bradac09-Bullet} panel shown in
Figure~\ref{fig-bullet}.  Furthermore, \citet{Randall2008}
explicitly analyzed the offset between galaxy and mass centroids and
found the mass to be $25\pm 29$ to the east (for comparison, the H15
offset is just over 100 kpc to the northwest). We adopt the
\citet{Randall2008} offset.

Each of these updated lensing offsets must be corrected for the gas
mass contribution as described at the start of this section.  This is
the system least likely to need such a correction, because the lensing
contours are so clearly separated from the bulk of the gas.
Nevertheless, to avoid any possible bias toward SIDM we apply the mean
H15 correction to each substructure.  This correction leads to a final
offset of $\delta_{SI}=-4.3$ kpc in the East, and $21$ kpc in the West.


\subsection{MACS J2243.3-0935}

\begin{figure}
\centerline{\includegraphics[scale=0.65]{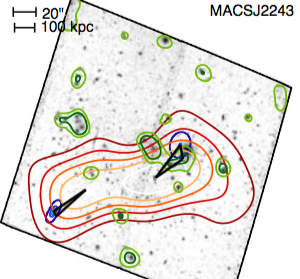}}
\vskip5mm
\centerline{\includegraphics[scale=0.65]{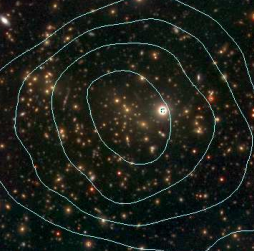}}
\caption{{\it Top:} View of MACS J2243 from H15.  {\it Bottom:} mass
  contours from \citet{WtGI}.}\label{fig-macs2243}
\end{figure}

Both subclusters in this system have substantial weight (4.9\% for the
east and 3.4\% for the west), so we treat them together.  The bottom
panel in Figure~\ref{fig-macs2243} is from \citet{WtGI}, who performed
a careful weak lensing analysis supported by 10 bands of photometry.
In addition, \citet{MACS2243-Schirmer} analyzed a larger field
including this system, using five-band CFHT/Megaprime imaging in good
seeing.  They used photometric redshifts to support source selection
and member galaxy identification, and further used 190 spectroscopic
redshifts to finely calibrate the photometry and obtain $\sim 0.03$
rms redshift uncertainty.  Their maps cover a much broader field at
lower resolution and so are not presented here, but they confirm the
\citet{WtGI} picture of a single smooth, round, high S/N mass
concentration centered on the galaxy concentration that appears at the
center of the bottom panel of Figure~\ref{fig-macs2243}.

The top and bottom panels of Figure~\ref{fig-macs2243} look rather
different at first, so we advise the reader to focus first on the
bright star that dominates the bottom panel.  This star is also the
brightest (most black) object in the upper panel, but there much of it
is obscured by the blue, green, and red contours that run over it.
There are two dense concentrations of galaxies, one immediately to the
left (east) of this star and the other farther east.  These galaxy
concentrations, with heavy green contours in the H15 map, are useful
points of reference when comparing the maps, but in fact neither is
used by H15 as a substructure.  We consider the H15 substructures
individually below.

{\it East:} the triangle representing this substructure is seen at the
lower left of the H15 image. However, there is no convincing
concentration of galaxies at this position in either panel of
Figure~\ref{fig-macs2243}, nor in
\citet{MACS2243-Schirmer}. Furthermore, neither \citet{WtGI} nor
\citet{MACS2243-Schirmer} find a lensing peak here.  The absence of
confirmation in more extensive data sets suggests that the H15 lensing
peak is spurious.  We omit this substructure from the remaining
analysis.

{\it West:} the color image from \citet{WtGI} demonstrates that the
H15 luminosity peak just south of the bright star cannot be due to
{\it galaxy} luminosity.  Although H15 implemented an algorithm for
removing stars before smoothing the visible light distribution, the
most plausible explanation for this H15 luminosity peak is residual
light from the exceptionally bright star.  The green contours to the
east of the bright star in the H15 panel represent a more valid galaxy
luminosity position.  This yields $\delta_{SI}\approx 0$ because the
galaxy-lensing offset is perpendicular to the galaxy-gas offset.  The
following paragraph contains more details for those with a special
interest; most readers are encouraged to skip to the next subsection.

For completeness, we note that the lensing maps also disagree: the
\citet{WtGI} lensing map is clearly centered on the galaxies rather
than on the blue H15 contour above the bright star.  The \citet{WtGI}
map \citep[confirmed at lower resolution by][]{MACS2243-Schirmer}
suggests that the offset between lensing and galaxy luminosity is
approximately zero.  In other words, with the correct luminosity
position the offset is roughly zero regardless of the lensing map we
adopt. The associated uncertainty is difficult to quantify from the
information in \citet{WtGI} and \citet{MACS2243-Schirmer}, but the
default H15 value of 60 kpc is a reasonable estimate.  The gas mass
correction is small compared to this uncertainty, but we apply it
nevertheless to avoid bias in the ensemble.  This yields
$\delta_{SI}=-4.3$ kpc after gas mass correction.

\subsection{ZwCl 1358+62 (East)}

\begin{figure}
\centerline{\includegraphics[scale=0.65]{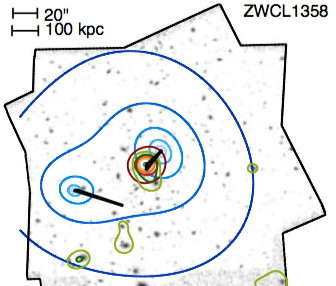}}
\centerline{\includegraphics[scale=1.5,angle=27.5]{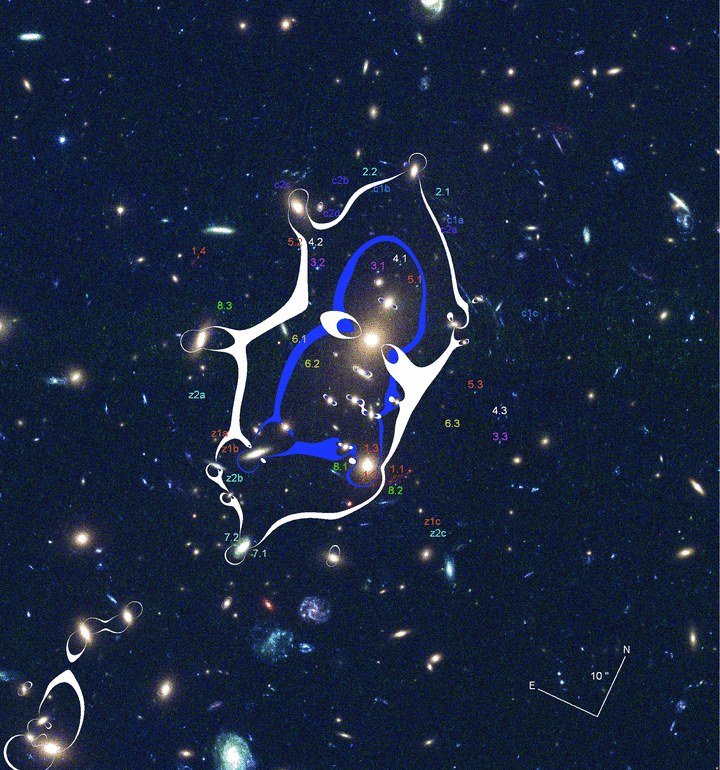}}
\caption{{\it Top:} View of ZwCl 1358 from H15.  {\it Bottom:} lensing
  critical curves from \citet{Zitrin11}, showing a secondary mass peak
  to the south rather than to the east as shown by
  H15.}\label{fig-ZwCl1358}
\end{figure}

\citet{Zitrin11} performed a strong-lensing analysis of this cluster
using deep six-band HST/ACS imaging.  They found 23 images of eight
different sources to support the construction of a mass model.  The
resulting critical curves are shown on top of a color ACS image in
Figure~3 of \citet{Zitrin11}, repeated here as the bottom panel of
Figure~\ref{fig-ZwCl1358}.  The H15 panel shows X-ray contours
strongly peaked on the BCG; this system also has a beautiful low
surface brightness X-ray tail, not visible here, extending along with
the galaxies to the south-southeast.\footnote{See the H15 press
  release image at
  \url{http://chandra.harvard.edu/photo/2015/dark/dark_zwcl1358.jpg}.}
This strongly suggests a merger axis along a south-southeast
direction.  The \citet{Zitrin11} strong-lensing mass reconstruction
matches the mass distribution one would expect in this situation:
elongated toward the south and with a secondary peak corresponding to
the southern galaxies.  In this context, the H15 weak-lensing finding
of a large mass to the east, and none to the south, is difficult to
explain.

The putative H15 mass is just off the eastern edge of the
\citet{Zitrin11} map, so there remains some possibility that the H15
mass peak is real. If so, there is essentially no associated gas peak.
The X-ray tail has nearly uniform (low) surface brightness, but H15
apparently identified a local maximum and automatically matched it to
a peak in their mass map much further to the east.  There is no
plausible connection between the X-ray tail and this putative
substructure in the east, as the X-ray morphology indicates a merger
along an axis from the south-southeast to the north-northwest.
Finally, the associated luminosity peak is not convincing either,
consisting of one or two galaxies.  We recommend omitting this
substructure from the sample.

Observant readers may notice that the \citet{Zitrin11} and H15 results
also disagree in the main part of the cluster, near the X-ray peak.
We do not examine this further because this substructure carries only
0.006\% of the total weight of the sample due to the small offset
between X-rays and galaxies.  Also, readers wishing to search the
literature on this cluster should be aware that it has several names,
including but not limited to ZwCl 1358.1+6245, MS 1358.4+6245, MACS
J1359.8+6231, and RXC J1359.8+6231.

\subsection{MACS J0025.4-1222 (West)}\label{sec-MACS0025}

\begin{figure}
\centerline{\includegraphics[scale=0.65]{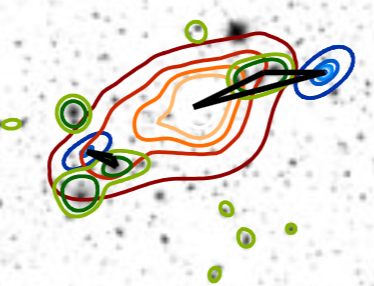}}
\centerline{\includegraphics[scale=0.58]{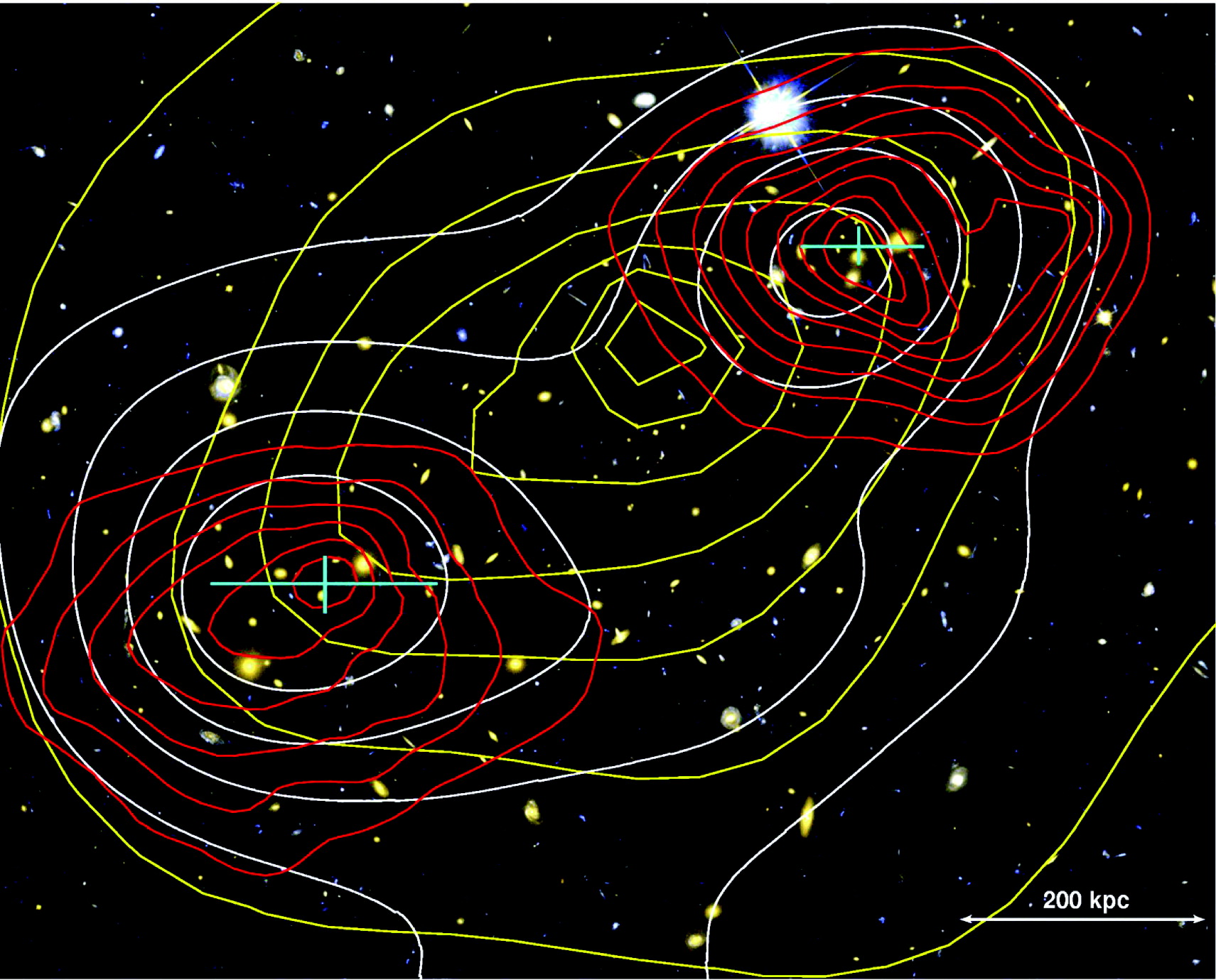}}
\caption{{\it Top:} View of MACS J0025 from H15.  {\it Bottom:}
  lensing (red), X-ray (yellow), and $I$-band light (white) contours from
  \citet{Bradac08-MACS0025}. Crosses show $1\sigma$ uncertainties on
    the mass positions.}\label{fig-MACS0025}
\end{figure}

Figure~\ref{fig-MACS0025} compares the H15 map with that of
\citet{Bradac08-MACS0025}, who used strong and weak lensing supported
by deep three-band HST and five-band Subaru imaging.  Here we are
concerned only with the western substructure, on the right of each
panel.  The \citet{Bradac08-MACS0025} mass peak (red contours) matches
the position of the luminosity contours in either panel, but not the
H15 mass contours.  In fact, the cyan cross in the
\citet{Bradac08-MACS0025} panel gives the $1\sigma$ error bar on the
mass position, showing that the H15 position is excluded at about the
$3\sigma$ level.  However, \citet{Bradac08-MACS0025} still find the
mass to be slightly ahead of the galaxies, in large part because their
galaxy position is a bit further back than the H15 position.  

According to Table~2 of \citet{Bradac08-MACS0025}, the galaxy-mass
offset is $-4.4$ arcsec (-29 kpc).  With the mean H15 correction for
gas mass, the final value for $\delta_{SI}$ is $-33$ kpc.  This is a
substantial change from the H15 value of $-151$ kpc.

A recent strong lensing analysis \citep{Cibirka18} lends further
  support to the \citet{Bradac08-MACS0025} results. This is the only
  substructure in our review where strong lensing is the major factor
  in determining an offset, so it is worth remarking that strong and
  weak lensing may correctly yield different positions if the region
  of highest surface mass density (probed by strong lensing) is not
  centered on the larger region of lower surface mass density (probed
  by weak lensing). In fact, SIDM simulations
  \citep{RobertsonBullet2017,Kim2016,Kahlhoefer14} show a low-density
  bridge forming between the two peaks after first pericenter in a
  bimodal merger. This shifts the centroid but not the peak, which
  stays with the galaxies. Thus, a high-confidence weak lensing
  position is in principle preferable to a strong lensing position
  when probing SIDM.  However, the outermost \citet{Bradac08-MACS0025}
  contours are indeed driven by (multiband) weak lensing, and the H15
  (single-band) weak lensing position is still 110 kpc from their
  geometric center.  Furthermore, the outer \citet{Bradac08-MACS0025}
  contours extend in a direction opposite to that expected from the
  bridge argument. We thus do not find the H15 position to be
  supported by the bridge argument. In general, strong+weak lensing
  position potentially disfavor SIDM by missing possible bridge
  effects, so we use the \citet{Bradac08-MACS0025} position to build
  the case that we need not make analysis choices favorable to SIDM to
  show that the H15 upper limits on SIDM must be revised substantially
  upward. However, one could also argue the opposite because the weak
  lensing contours extend in the ``wrong'' direction. For that reason,
  in Section~\ref{sec-results} we assess the impact of this particular
  choice on the final result; it is small.

\subsection{ACTCL J0102-4915 (El Gordo, North)} 

\begin{figure}
\centerline{\includegraphics[scale=0.475]{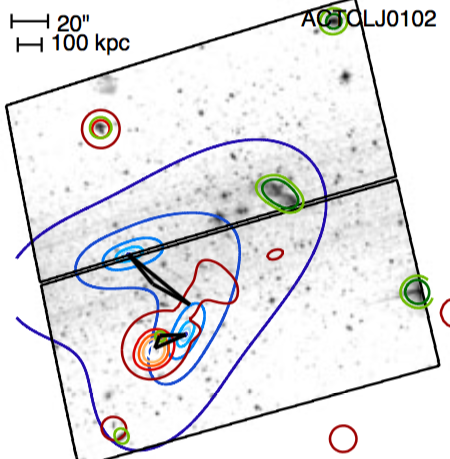}}
\centerline{\includegraphics[scale=0.6]{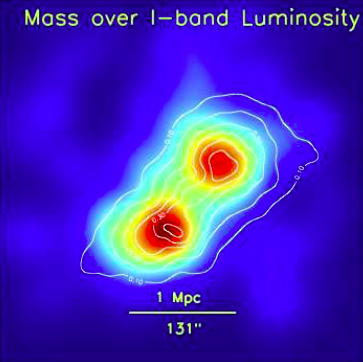}}
\caption{{\it Top:} H15 analysis of ACTCL J0102 (El Gordo). {\it
    Bottom:} lensing contours (white) and galaxy luminosity density
  (colormap) from \citet{JeeGordo}.  There is severe disagreement over
  the location of the northern lensing peak.  The \citet{JeeGordo}
  location is coincident with the galaxies, while the H15 location
  appears to be an artifact of the ACS chip gap.}\label{fig-gordo}
\end{figure}

Figure~\ref{fig-gordo} compares the H15 analysis with that of
\citet{JeeGordo}, who used four-band imaging in conjunction with the
photometric redshift catalog of \citet{Menanteau12} and 89
spectroscopic redshifts \citep{Sifon13} to obtain clean background
source selection---a particularly important issue when the lens is
itself at fairly high redshift ($z=0.87$).  \citet{JeeGordo} found two
mass concentrations roughly coincident with the northwest and
southeast galaxy concentrations, and found no other mass
concentrations in the area.  H15 differ starkly in finding a mass
concentration not at the northwest galaxy location, but $\sim 700$ kpc
away.  The H15 location is remarkably coincident with the gap between
detectors in the Advanced Camera for Surveys (ACS) data used by H15.
We suggest that this is a spurious lensing peak related to the
difficulty of cleaning data in or near the gap. We find further
support for this suggestion in the strong lensing analysis of
\citet{Zitrin13}, whose critical curves are consistent with the mass
distribution of \citet{JeeGordo} but not that of H15.

Other aspects of the northern H15 substructure are problematic as
well. The H15 galaxy position does not look like a galaxy overdensity
in color images, and even the H15 figure panel lacks green contours at
this location (the middle vertex of the triangle).  Furthermore, the
H15 triangle suggests a southwest-northeast merger axis, but an
overwhelming variety of other evidence \citep[all of the above-cited
papers plus the radio relics presented in][]{Lindner13} supports a
southeast-northwest merger axis; nowhere in the extensive literature
on El Gordo is there any evidence for a southwest-northeast merger
axis.

In summary, H15 have incorrectly characterized the northern
substructure. We may be able to infer $\delta_{SI}$ from
\citet{JeeGordo} and/or \citet{Zitrin13}.  However, there is no gas
peak associated with the northwest subcluster.  The detailed X-ray map
of \citet{Menanteau12} shows a large region of tenuous gas, but no
identifiable peak.  In this situation, even a peak-agnostic algorithm
such as a centroid would have a very large associated uncertainty,
implying that this substructure would have little influence on the
ensemble.  Therefore, we recommend omitting this substructure from the
ensemble.

\subsection{MACS J0417.5-1154 (North)}

\begin{figure}
\centerline{\includegraphics[scale=0.475]{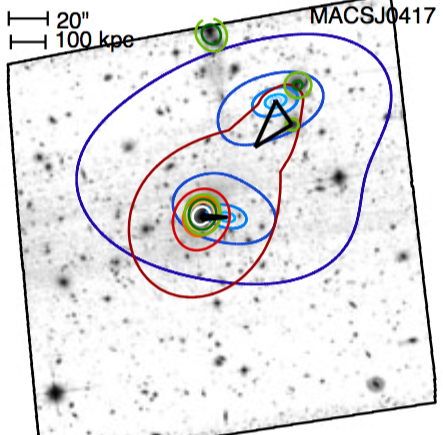}}
\caption{H15 analysis of MACS J0417. Only the northern subcluster is
  highly weighted and considered here. The X-ray morphology there
  consists of a long ridge rather than a peak. The luminosity peak at
  the top is a star and should be disregarded.}\label{fig-0417}
\end{figure}

\citet{WtGI} performed a weak lensing analysis of this cluster using
three-band Subaru imaging.  Their lensing map has rather low
resolution and so is not presented here, but it does show the same
southeast-northwest axis as the H15 lensing map
(Figure~\ref{fig-0417}). The galaxy distribution and X-ray morphology
follow the same axis, so there is no reason to doubt the H15 lensing
map.

Nevertheless, this substructure is worth discussing to illustrate some
ambiguities facing next-generation analyses of this sort.  First, as
suggested by the H15 X-ray contours, the X-ray morphology
\citep{MannEbeling2012} shows a long ridge to the northwest with no
peak other than the main peak identified with the southern subcluster.
H15 presumably identified a minor local maximum as the subcluster gas
location, but this is a somewhat arbitrary location along a long
smooth ridge.  Without a clear X-ray peak along this ridge, there is
great uncertainty in the galaxy-gas vector and therefore in the weight
this substructure should receive as well as in the projection of the
DM-galaxy vector onto the galaxy-gas vector.  Second, the H15 DM peak
appears midway between two luminosity peaks.  The H15 matching
algorithm chose the luminosity peak nearer the putative gas peak,
yielding $\delta_{SI}=2$ kpc after projection onto the galaxy-gas
vector).  Automatically matching the nearest peak may introduce a
bias: if the brighter luminosity peak had been adopted, this substructure
would have different implications for SIDM, with
$\delta_{SI}\approx 50$ kpc over a longer ($\sim 200$ kpc) galaxy-gas
baseline providing a great deal of weight.  Or, with more smoothing of
the light, the luminosity location would be intermediate and the
implication for SIDM would be intermediate. This exposes the need to
develop methods less sensitive to smoothing scale, or at least an
objective way to optimize the smoothing scale in each system.

\subsection{ZwCl 1234+2916 (West)}

\begin{figure}
\centerline{\includegraphics[scale=0.8]{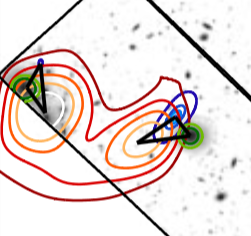}}
\centerline{\includegraphics[scale=0.5]{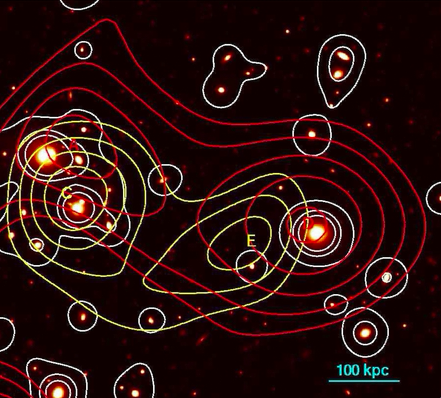}}
\caption{{\it Top:} H15 analysis of ZwCl 1234. {\it Bottom:} lensing 
  (red), galaxy (white), and X-ray (yellow) contours from 
  \citet{Dahle13}.}\label{fig-1234}
\end{figure}

This cluster illustrates the level of agreement we would expect for
independent investigations using overlapping
data. Figure~\ref{fig-1234} compares the H15 analysis with that of
\citet{Dahle13}, who used two-band VLT imaging as well as the
extremely deep ACS imaging in a third band (the data used by H15).
The \citet{Dahle13} lensing contours (in red) agree with H15 in
showing each mass concentrations slightly to the north and, in the
east-west direction, slightly closer to the center of the system
compared to the corresponding galaxies. Here we are concerned with the
western substructure.  Although H15 put the mass concentration
slightly farther north than do \citet{Dahle13}, this displacement is
nearly perpendicular to the merger axis and so has little effect on
$\delta_{SI}$.  \citet{Dahle13} do not list a value of $\delta_{SI}$,
but measurements of their map yield $\approx 10$ kpc, close to the H15
value of 28 kpc.  Although the agreement is not perfect, this level of
variation is to be expected in independent analyses.  We retain the
H15 offset.

\subsection{Summary\label{subsec-summary}}

Table~\ref{tab-changes} summarizes the updates we recommend after
reviewing the best available evidence from the literature.  In four
cases the literature is consistent with the offset ($\delta_{SI}$)
measured by H15; in five cases there is compelling evidence that the
offset differs from that of H15; and in seven cases the H15
substructure either does not exist or does not support a clear
association between a gas peak, a mass peak, and a galaxy peak. 
Although few of the H15 offsets are retained, this set of changes is
the minimum necessary to bring the H15 offset catalog in line with the
literature.  In the next section we quantify the impact of these
changes on the dark matter inference.

\begin{table}
\caption{Cluster Weights and Galaxy-DM Offsets}
\begin{tabular}{rrcll} 
Weight & \multicolumn{2}{c}{$\delta_{\rm SI}$ (kpc)}&&\\
\cline{2-3}
(\%) & H15  & Literature\tablenotemark{a}  & Name & \S \\ \tableline
16.7 & 66 & Omit (M11) &Abell 2744 (west)&3.1\\ 
15.0 & -19 & 80 (D13) &DLSCL J0916 (south) &3.2\\ 
10.1 & 36 & No change &Abell 520-1 &3.3\\ 
6.1 & 40 & -4 (B09) &Bullet (east) &3.4\\ 
4.9 & 4 & Omit (L14, S11) &MACS J2243 (east) &3.5\\ 
4.7 & -7 & Omit (Z11) &ZwCl 1358 (east) &3.6\\ 
4.7 & 81 & Omit (J14a) &Abell 520-2 &3.3\\ 
4.1 & -151 & -33 (B08)&MACS J0025 (west) &3.7\\ 
3.8 & -32 & 21 (B09,R08) &Bullet (west) &3.4\\ 
3.4 & -26 & -4 (L14) &MACS J2243 (west) &3.5\\ 
2.9 & 22 & Omit (J14a) &Abell 520-5 &3.3\\ 
2.4 & -150 & Omit (J14b) &ACTCL J0102 (north) &3.8\\ 
2.1 & 2 & No change &MACS J0417 (north)&3.9\\ 
1.8 & 28 & No change &ZwCl 1234 (west)&3.10\\ 
1.7 & 84 & Omit (C12,J14a) &Abell 520-3 &3.3\\ 
1.6 & -22 & No change &Abell 520-4 &3.3\\
\tableline \end{tabular}
\tablenotetext{1}{References: B08 \citep{Bradac08-MACS0025};
B09 \citep{Bradac09-Bullet};
C12 \citep{Clowe12-A520};
D13 \citep{DawsonPhDT};
J14a \citep{Jee14-A520};
 J14b \citet{JeeGordo};
L14 \citep{WtGI};
M11 \citep{Merten11};
R08 \citep{Randall2008};
S11 \citep{MACS2243-Schirmer},
Z11 \citep{Zitrin11}.
}\label{tab-changes}
\end{table}


\section{Dark matter inference}\label{sec-results}

We begin with a brief recap of the H15 prescription for inferring
$\sigmaDM$ from the offset catalog. H15 approximate the likelihood for
each $\beta_i$ as a Gaussian, as described in
Section~\ref{sec-technique}.  They multiply these likelihoods to find
a likelihood for $\langle \beta \rangle$ and then transform this into
a likelihood for $\sigmaDM$ according to the relation
$\sigmaDM = -\sigma_* \ln(1-\beta)$, where $\sigma_*$ is the
characteristic cross-section (per unit mass) at which a halo becomes
optically thick.  They choose a central value of $\sigma_* = 6.5$
cm$^2$/g and ``analytically marginalize'' over the range
$3.5<= \sigma_* <= 9.5$ by quadrature addition to the second moments
of the $\sigmaDM$ likelihood.

We prefer to marginalize numerically so that any non-Gaussian features
can be preserved. We create a grid of models in the
$(\sigmaDM, \sigma_{*})$ parameter space, covering the region
$(0-5,3.5-9.5)$.  For each point in the grid we multiply the $\beta_i$
likelihoods as determined by the H15 prescription.  We then
marginalize over the $\sigma_{*}$ axis to obtain a likelihood.

\begin{figure}
\centerline{\includegraphics[width=3.5in]{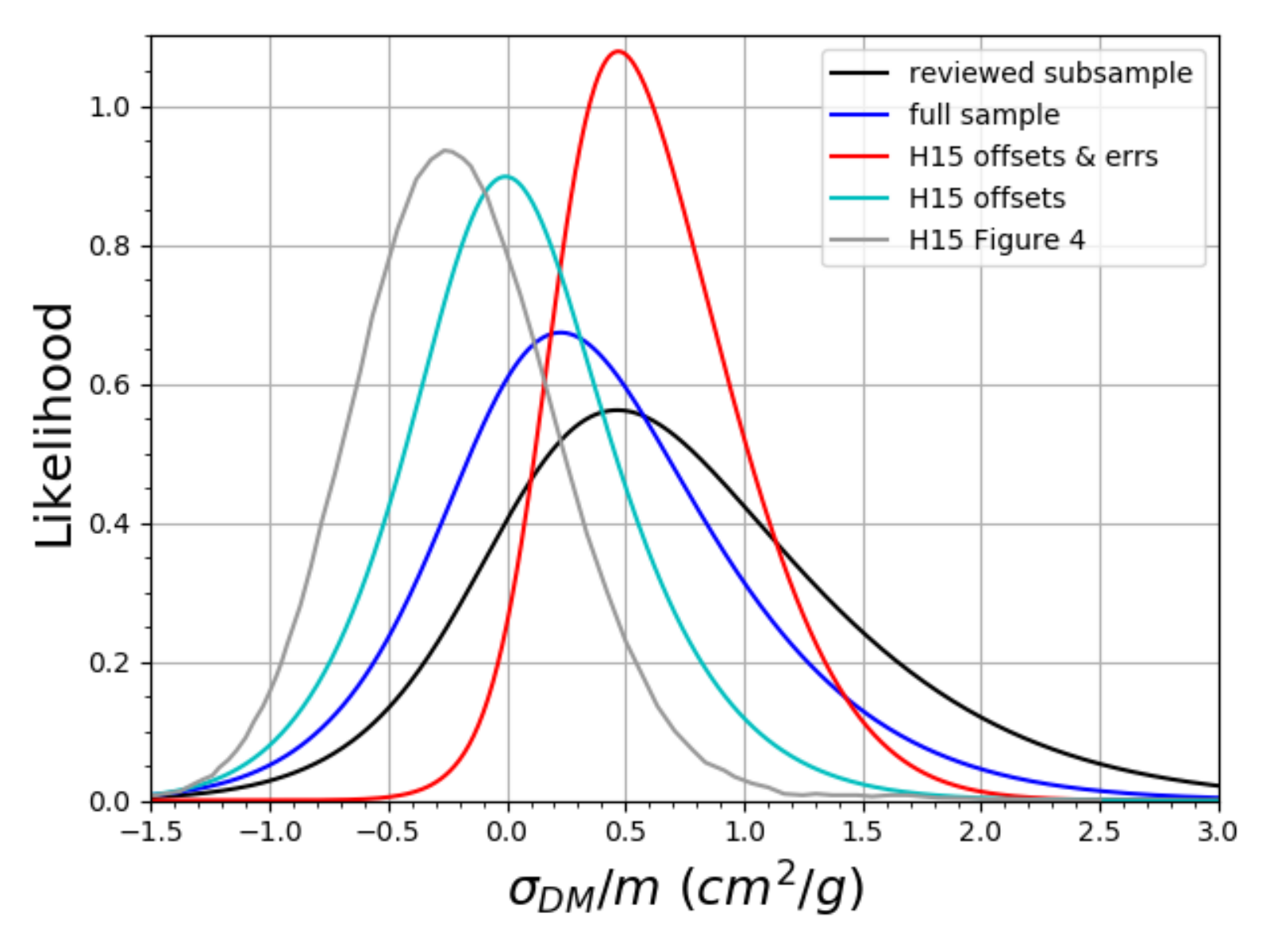}}
\caption{Likelihoods, normalized to unit area, for variations on the
  cross-section analysis. Our marginalization using the H15 offsets
  and their adopted 60 kpc uncertainty on each offset (cyan) agrees
  with the H15 result (gray) to well within $1\sigma$. The blue curve
  incorporates corrections to the offsets based on our literature
  review of the most highly weighted substructures; the black curve is
  based {\it only} on these reviewed substructures; and the red curve uses the
  H15 offsets {\it and} their individualized uncertainties as
  displayed in H15 Figure~S2.}\label{fig-pdf}
\end{figure}

Figure~\ref{fig-pdf} shows a digitized version of H15 Figure 4 (gray),
along with the result of our marginalization procedure (cyan) using
their offsets (as well as their nominal 60 kpc uncertainty on each
offset). The two results differ only slightly, with peak locations
shifted by $\approx 0.25$ cm$^2$/g, or $\approx 0.6$ times the
uncertainty given by H15. Correcting the most highly weighted offsets
based on our literature review, as listed in Table~\ref{tab-changes},
then yields the blue curve. Using {\it only} the offsets from the
vetted systems yields the black curve. The black peak is shifted from
the gray by about twice the uncertainty given by H15, still not highly
significant. For context, the red curve shows the shift that can
result from an even simpler variation on the H15 analysis: adopting
the individual offset uncertainties displayed in their Figure~S2
(rather than the uniform 60 kpc uncertainty they adopted in their
analysis) while retaining their offset measurements.  This also yields
a $\approx 2\sigma$ shift in the same direction.

Models with $\sigmaDM<0$ are not physical. To constrain physical
models, we apply a prior that is uniform for $\sigmaDM\ge 0$ and zero
otherwise.  Note that this will have little effect when using, say,
the red likelihood in Figure~\ref{fig-pdf}, because this likelihood
already nearly vanishes for $\sigmaDM<0$. Conversely, the prior will
have a substantial effect when using the gray likelihood, which peaks
in negative territory.  In the following, we show results with and
without the prior for completeness.

\begin{figure}
\centerline{\includegraphics[width=3.5in]{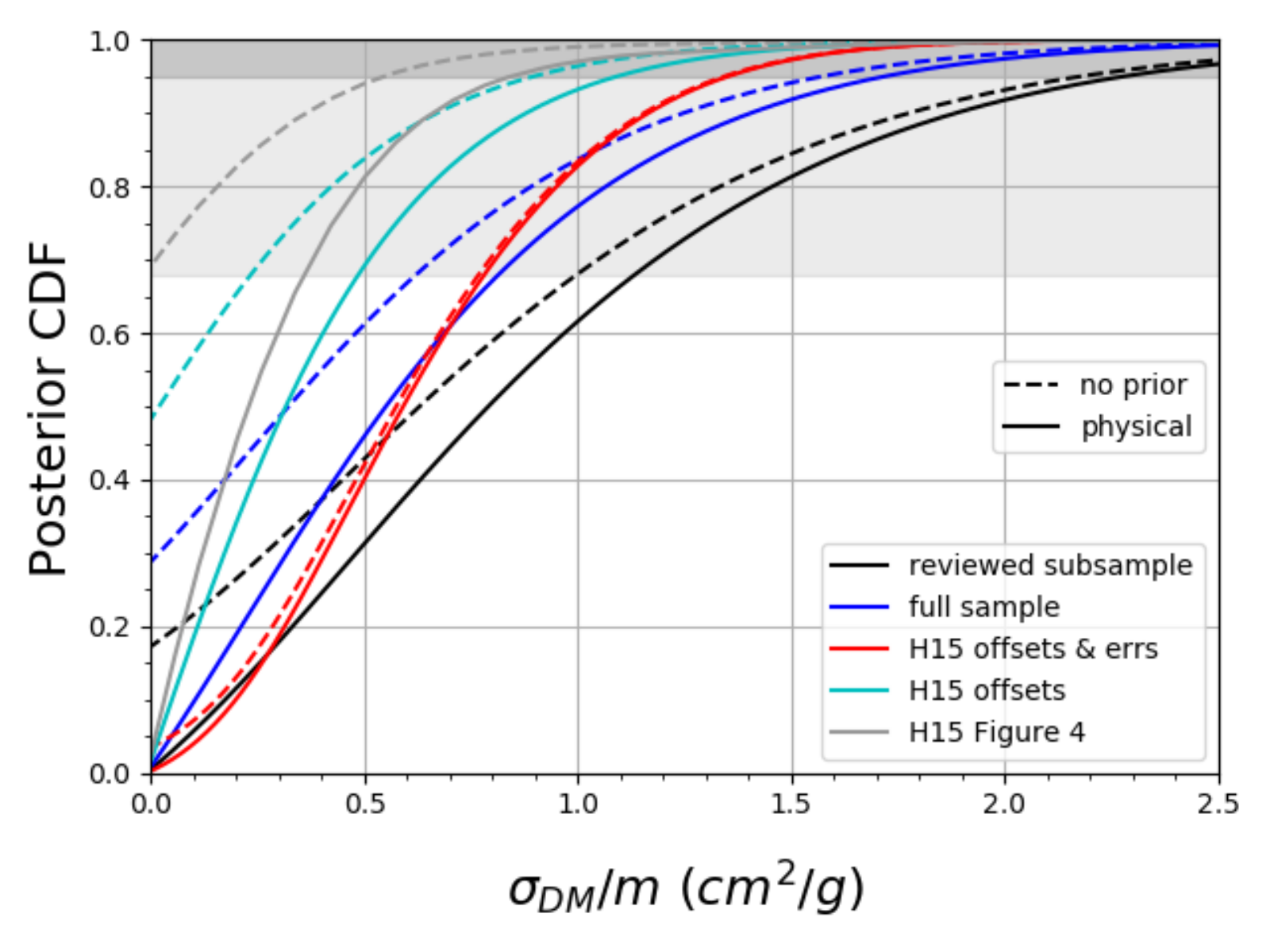}}
\caption{Constraints on $\sigmaDM$ with (solid) and without (dashed)
  the physical prior $\sigmaDM\ge 0$. All variations on the H15
  analysis lead to substantially relaxed upper limits.  The reviewed
  samples lead to 95\% upper limits of $\approx 1-2$ cm$^2$/g
  regardless of prior. The prior becomes more important when
  interpreting the H15 PDF, because most of its area is at negative
  cross section. Color codes are the same as in Figure~\ref{fig-pdf}.}\label{fig-cdf}
\end{figure}

After optional application of the prior, we integrate each posterior
into a cumulative distribution function (CDF), which allows the reader
to quickly read an upper limit at any desired confidence level.
Figure~\ref{fig-cdf} shows CDFs with (solid) and without (dashed) the
physical prior, with color coding to match Figure~\ref{fig-pdf}.  As
an example of how to read this figure, the dashed gray (H15) curve in
Figure~\ref{fig-cdf} indicates 70\% confidence that $\sigmaDM<0$ and
95\% confidence that $\sigmaDM<0.5$ cm$^2$/g with no prior, but the
solid curve indicates substantially less confidence in those limits
after applying the prior. In contrast, the red solid and
dashed curves in Figure~\ref{fig-cdf} are nearly identical; this means
that the constraint using the H15 offsets and uncertainties is nearly
unaffected by the prior, as predicted in the previous paragraph.  The
95\% confidence limits from our reviewed subsample, or the full
(corrected) sample, are only marginally affected by the prior.

Regardless of the prior, Figure~\ref{fig-cdf} reveals a dramatic
relaxation of upper limits when using the best available information
about each cluster: $\sigmaDM<1.71$ cm$^2$/g at 95\% confidence for
the full sample, or 2.27 cm$^2$/g for the reviewed subsample.  This is
driven by the data, not the prior: the upper limits change by 0.15
cm$^2$/g or less when dropping the prior for these data. On the other
hand, the strict upper limit quoted by H15 for their data is possible
{\it only if} the physical prior is discarded.

So far our discussion has considered any (nonnegative) value of
$\sigmaDM$ to be equally likely {\it a priori}. If instead we treat
$\sigmaDM=0$ (cold dark matter) as a null hypothesis, it is clear that
all likelihoods in Figure~\ref{fig-pdf} are consistent with this
hypothesis.

We tested the sensitivity of the analysis to additional variations.
As a reminder, our default analysis follows H15 in assigning 60 kpc
uncertainty to each value of $\delta_{SI}$---not because we endorse
this procedure, but to demonstrate that a {\it minimal} set of changes
to the H15 offsets and procedures yields substantially looser
constraints. In one variation, we used the uncertainties displayed by
H15 in their Figure S2 with our literature-based offsets.  In a second
variation, we inserted literature-based uncertainties where explicitly
available, i.e. for the Bullet Cluster West \citep[$25\pm29$
kpc,]{Randall2008}. In a third variation, we explored the sensitivity
to gas mass correction by adding the mean 4.3 kpc correction back to
each value of $\delta_{SI}$.  In all cases the constraint shifted by
substantially less than the difference between the black and blue
curves in Figure~\ref{fig-cdf}.

We also tested the sensitivity to individual updates.  The
  omission of Abell 2744 West lowered the most likely value of
  $\sigmaDM$ by 0.19 cm$^2$/g, and the update to DLSCL J0916 South
  raised it by 0.24 cm$^2$/g; other individual updates had at most
  half this effect.  This suggests that our scheme for estimating the
  weight of each subcluster in the ensemble analysis is effective, and
  that the most important subclusters have indeed been reviewed.  Note
  that the size of the proposed change matters as well as the weight
  of the subcluster: omitting the remarkably large negative offset of
  ACTCL J0102 (based on a spurious lensing position) had the third
  largest effect (a $+0.12$ cm$^2$/g shift), outstripping smaller
  adjustments to more highly weighted subclusters. Updating MACS J0025
  West from $-151$ to $-33$ kpc had the fourth-largest effect, a
  $+0.11$ cm$^2$/g shift. Thus, if one favors weak lensing contours as
  discussed in \S\ref{sec-MACS0025} and uses a slightly more negative
  offset for this substructure, the ensemble result would be lowered
  by a fraction of $0.11$ cm$^2$/g.

To summarize this section, when using the H15 offsets we find a
  most likely value of $\sigmaDM$ that is slightly higher than, but
  consistent with, the H15 value. After updating the most highly
  weighted offsets according to our literature review, the most likely
  value shifts further upward to $+0.23$ cm$^2$/g (blue curve in
  Figure~\ref{fig-pdf}).  The corresponding upper limit (at 95\%
  confidence) is 1.71 cm$^2$/g if one adopts a physical prior (as we
  recommend), or 1.57 cm$^2$/g if one does not.  These limits are
  substantially higher than found by H15, and increase further by
  about 0.5 cm$^2$/g if one uses {\it only} the reviewed subsample of
  highly weighted subclusters. Our higher upper limits are robust
  against many variations in the analysis.

\section{Summary and discussion}\label{sec-discuss}



Adopting the H15 methodology but with corrections to their offsets
based on the best available evidence in the literature and
marginalizing only over physical models, we find a 95\% confidence
upper limit on $\sigmaDM$ of $\lesssim 2$ cm$^2$/g, depending on the
whether one uses the high-quality subsample or the full sample.  In
other words, the H15 methodology does not yet support constraints
tighter than the 1.25 cm$^2$/g at 68\% confidence quoted by
\citet{Randall2008}.  In this discussion we first defend the
literature-review approach that led to this conclusion. Then, we
explain why the H15 sample selection should lead readers to use even
the revised constraints with caution.  Finally, we 
discuss future prospects.

We predict two types of concerns readers may have with
our approach:
\begin{itemize}
\item Bias: was evidence from the literature applied consistently
  without regard to the effect on the final result?  It is impossible
  to remain ignorant of the potential impact of a change in the offset
  when reviewing the H15 star-gas-DM geometry, because the geometry is
  so simple: if DM is self-interacting then it should be bracketed by
  the stars and the gas.  We have attempted to minimize this concern
  by systematically examining the highest-weight substructures rather
  than the most negative or most suspect offsets.  As {\it post facto}
  evidence that our review was not biased, we note that of the seven
  H15 substructures we omitted, five had positive offsets in H15, and
  only two had negative offsets.  By itself, this should bring
  $\sigmaDM$ down rather than up, barring complications such as the
  differing weights and sizes of the offsets.  Of the changes to
  $\delta_{SI}$ that we recommend, the majority do go in the direction
  of lifting $\sigmaDM$ from its H15 value, but the literature on
  these systems is so compelling that it speaks for itself.

\item Inhomogeneity: the H15 catalog was produced in a mostly uniform
  way and uses only ACS data for photometry and lensing, but our
  corrections are based on a variety of data and analyses from the
  literature.  While uniformity is a laudable goal, we do not believe
  it overrides the compelling evidence in the literature.  Indeed, we
  believe the concern for uniformity is a major reason some of the H15
  offsets are in error: single-band photometry is not sufficient to
  adequately characterize the mass and luminosity distribution of
  these systems.

\end{itemize}
These arguments suggest that any ill effects of bias and inhomogeneity
are likely to be smaller than the beneficial effect of using more
correct offsets.  

We now turn to concerns about sample selection.  The equation used by
H15 to relate $\beta$ to $\sigmaDM$ was developed assuming only
small ($\lesssim 30$ kpc) displacements between DM, galaxies and gas;
\citet{Harvey14} clearly shows how the analogy between galaxy and gas
restoring forces (and hence the inferred drag force) breaks down
quickly beyond 30 kpc displacements. Yet, many of the H15 offsets are
very far outside this regime.  Figure S2 of H15 readily shows that
only five of the 72 substructures have both $\delta_{SI}$ and
$\delta_{SG}$ within 30 kpc. Our updates have modestly reduced the
spread of $\delta_{SI}$ values, but many violations of this
approximation remain.  In fact, the ensemble result is {\it driven} by
substructures that violate the approximation, because the weight of
each substructure is approximately proportional to $\delta_{SG}^2$
(Equation~\ref{eqn-3}).

The analogy between gas and self-interacting DM can break in other
ways as well, for example if the gas is completely stripped at
pericenter crossing.  The analogy also breaks after the drag force (if
any) subsides, for example well after pericenter as the subcluster
travels to regions of lower and lower density. Subsidence of the drag
force allows the gas and galaxies to each fall back to, and {\it
  through}, the DM.  Recent SIDM simulations by \citet{Kim2016}
clearly show the sign of the DM-galaxy offset changing as this
happens.  Separately, it has also been seen with gas in an effect
known as the ram pressure slingshot: after ram pressure pulls the gas
back from the center of the subcluster potential, gravity slings it
forward \citep{Markevitch2007,Mathis2005,Hallman04, Ng2015}.  Because
the maximum galaxy-DM displacement is smaller than the maximum gas-DM
displacement, the two components cannot fall through the DM on the
same timescale.  Thus the ratio of displacements $\beta$ cannot remain
constant over time, and could even change sign at times. Averaging
over a sample without regard to merger phase would then bias the
inferred cross section low.  We therefore suggest caution in
interpreting even the revised constraints.

Although this conclusion is disappointing for current constraints, it
does suggest that constraints from merging clusters could be tightened
with a closer analysis of key systems including modeling the merger
phase. We suggest that post-pericenter systems should be modeled with
hydrodynamic simulations \citep[e.g.,][]{RobertsonBullet2017}  rather
than with an analytical formalism.  Other aspects of cluster infall
are also potentially competitive \citep{Kahlhoefer15}.


Cluster mergers can also be used to test SIDM effects that cannot be
modeled with a drag force; as noted by \citet{Kahlhoefer14}, the
drag-force model maps well to frequent interactions with low momentum
transfer, as in a long-range force.  Infrequent interactions with
large momentum transfer (as in hard scattering), in contrast, can
eject particles from the cluster and is potentially observable as a
decrease in the mass-to-light ratio; \citet{Randall2008} used the
Bullet Cluster to constrain these models with an upper limit of
$\sigmaDM < 0.7$ cm$^2$/g (68\% confidence). This gives clusters
a purpose beyond constraining the cross section at high velocity: if
dark matter does interact with itself, comparing the two types of
cluster constraints could point the way to a particle model of the
interactions.

We close with a simple message: single-band imaging is
  insufficient to properly identify mass and light peaks. H15 argue
  that single-band imaging suffices for a large sample because errors
  will average out.  We showed, however, that some substructures have
  much more weight than others in the ensemble, and that single-band
  imaging enables errors on individual substructures to be quite
  large, up to hundreds of kpc.  As a result, the errors cannot be
  counted on to cancel over the H15 sample.

\acknowledgments We thank Maru\v{s}a Brada\v{c}, James Bullock, Bill
Forman, James Jee, Felix Kahlhoefer, Manoj Kaplinghat, and Reinout van
Weeren for helpful discussions, as well as the anonymous referee for
useful feedback. We especially thank David Harvey for being completely
open with his code and his data, patiently answering our questions,
and reviewing a complete draft.  DW and NG were supported by NSF grant
1518246. Part of this work performed under the auspices of the
U.S. DOE by LLNL under Contract DE-AC52-07NA27344.

\bibliographystyle{apj}
\bibliography{ms}

\end{document}